\documentclass[twocolumn,aps,showpacs,floatfix,prc]{revtex4}
\usepackage[dvips]{epsfig}
\begin{document}
%\draft

\title{ Hindrance in heavy-ion fusion for lighter systems of astrophysical interest }

\author{ Vinay Singh$^{1}$, Joydev Lahiri$^{1}$, Partha Roy Chowdhury$^{2}$ and D. N. Basu$^{1}$}

\affiliation{$^1$Variable Energy Cyclotron Centre, 1/AF Bidhan Nagar, Kolkata 700064, INDIA}
\affiliation{ $^2$ Chandrakona Vidyasagar Mahavidyalaya, Chandrakona Town, Paschim Medinipur, West Bengal 721201, India}

\email[E-mail 1: ]{vsingh@vecc.gov.in}
\email[E-mail 2: ]{joy@vecc.gov.in}
\email[E-mail 3: ]{royc.partha@gmail.com}
\email[E-mail 4: ]{dnb@vecc.gov.in} 

\date{\today }

\begin{abstract}

    The hindrance in fusion of heavy-ion reactions crops up in the region of extreme sub-barrier energies. This phenomenon can be effectively analyzed using a simple diffused barrier formula derived assuming a Gaussian distribution of fusion barrier heights. Folding the Gaussian barrier distribution with the classical expression for the fusion cross section for a fixed barrier, the fusion cross section is obtained. The energy dependence of the fusion cross section provides good description to the existing data on sub-barrier heavy-ion fusion for lighter systems of astrophysical interest. Using this simple formula, an analysis has been presented from $^{16}$O + $^{18}$O to $^{12}$C + $^{198}$Pt, all of which were measured down to $<$ 10 $\mu$b. The agreement of the present analysis with the measured values is better than those calculated even from the sophisticated coupled channels calculations. The relatively smooth variation of the three parameters of this formula implies that it may be exploited to estimate the excitation function or to extrapolate cross sections for pairs of interacting nuclei which are yet to be measured. Possible extensions of the present methodology and its limitations have also been discussed.

\end{abstract}

\maketitle

\noindent
\section{Introduction}
\label{section1}

    The phenomenon of fusion hindrance in heavy-ion fusion reactions for lighter systems may have important consequences on the nuclear processes taking place in astrophysical scenarios. The hindrance can seriously affect the energy generation by heavy-ion fusion reactions occurring in the region of extreme sub-barrier energies, encompassing reactions involving lighter systems, such as the reactions occurring in the stage of carbon and oxygen burning in heavy stars \cite{Ro91,Da98,Wa97} and their evolution and elemental abundances. It is well known that fusion excitation functions cannot be satisfactorily explained assuming penetration through a single, well-defined barrier in the total potential energy of a colliding nucleus-nucleus system. In order to reproduce shapes of the fusion excitation functions, especially at low near-threshold energies, it is necessary to assume coexistence of different barriers, a situation that is naturally accounted for in the description of fusion reactions in terms of coupled channel calculations involving coupling to various collective states \cite{We91,Ste95,Ti98,Tr01,St06,St07}.

    The aim of the present work is to obtain the nuclear fusion cross sections for heavy-ion fusion for lighter systems of astrophysical interest. The phenomenological description of the fusion excitation functions is achieved by assuming a Gaussian shape of the barrier distribution treating the mean barrier and its variance as free parameters and folding it with the classical expression for the fusion cross section for a fixed barrier with the distance corresponding to the location of the interaction barrier as another free parameter. The free parameters are then determined individually for each of the reactions by comparing the predicted fusion excitation function with experimental data. The energy dependence of the fusion cross section, thus obtained, provides good description to the existing data on sub-barrier fusion and capture excitation functions for lighter heavy-ion systems of astrophysical interest.

\noindent
\section{The fusion barrier distribution}
\label{section2}

    It is well known that the energy dependence of the fusion cross sections can not be well estimated assuming simply the penetration through a well-defined barrier in one-dimensional potential of a colliding nucleus-nucleus system. The heavy-ion fusion cross sections require interpretation \cite{Ro91} in terms of a distribution of potential barriers. The smoothening due to the quantum mechanical barrier penetration replaces set of discrete barriers by an effective continuous distribution. In order to reproduce shapes of experimentally observed fusion excitation functions, particularly at low, near-threshold energies, it is necessary to assume a distribution of the fusion barrier heights, the effect that results from the coupling to other than relative distance degrees of freedom. This is naturally achieved in coupled-channel calculations, involving the coupling to the lowest collective states in both colliding nuclei. The structure effects in the barrier distributions are neglected in the present work and for the distribution of the fusion barrier heights, a Gaussian shape for the barrier distribution $D(B)$ is assumed \cite{Wi04}. The barrier distribution is, therefore, given by

\begin{equation}
 D(B)=\frac{1}{\sqrt{2\pi}\sigma_B}\exp\Big[-\frac{(B-B_0)^2}{2\sigma_B^2}\Big] 
\label{seqn1}
\end{equation}
\noindent
where the two parameters, the mean barrier $B_0$ and the distribution width $\sigma_B$, to be determined individually for each reaction.

\begin{figure}[t]
\vspace{0.8cm}
\eject\centerline{\epsfig{file=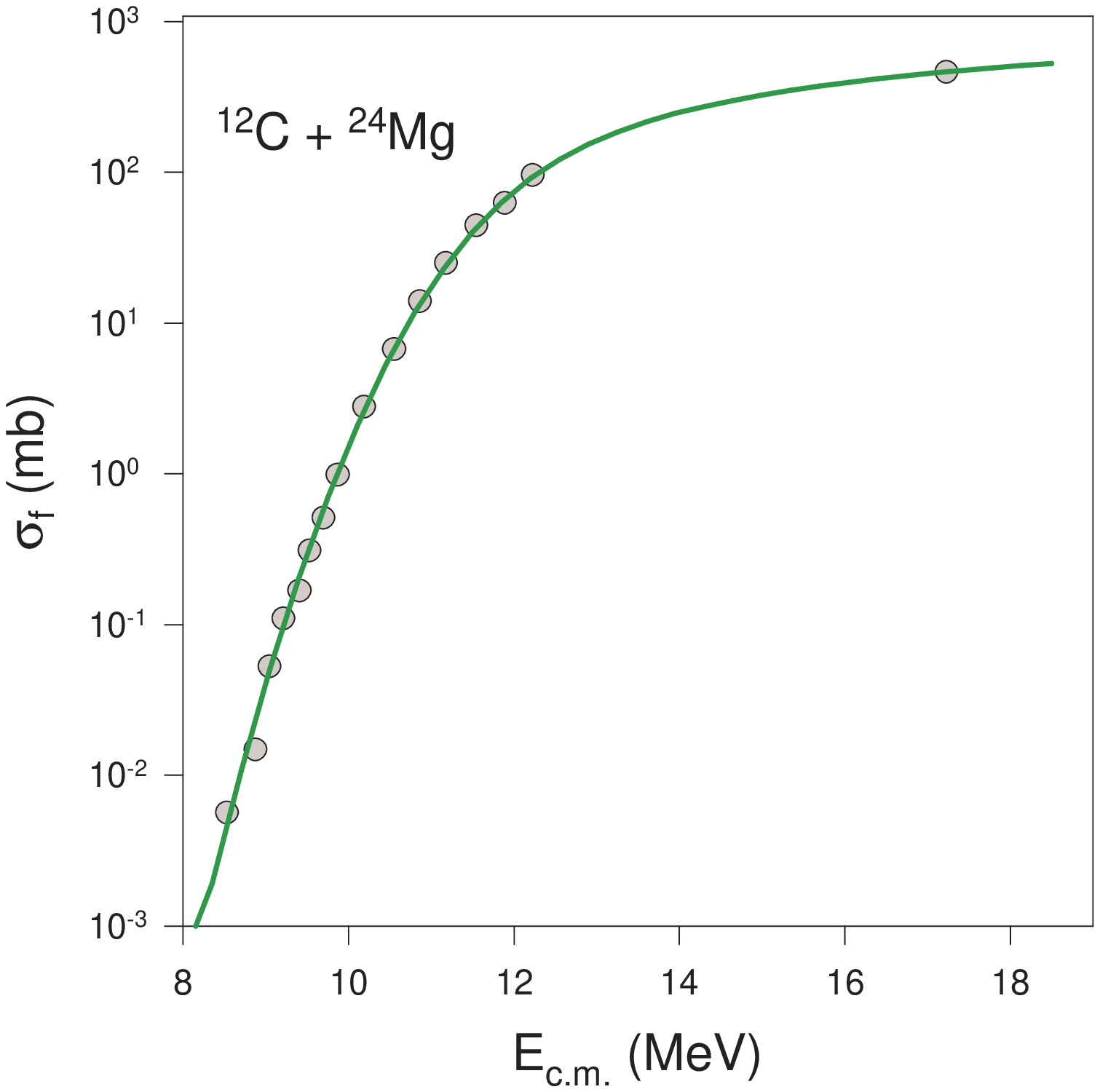,height=8cm,width=8cm}}
\caption
{Plots of the analytical estimates (solid line) and the measured values (full circles) of the capture excitation functions for $^{12}$C+$^{24}$Mg.}
\label{fig1}
\vspace{0.0cm}
\end{figure}
\noindent 

\begin{figure}[t]
\vspace{0.8cm}
\eject\centerline{\epsfig{file=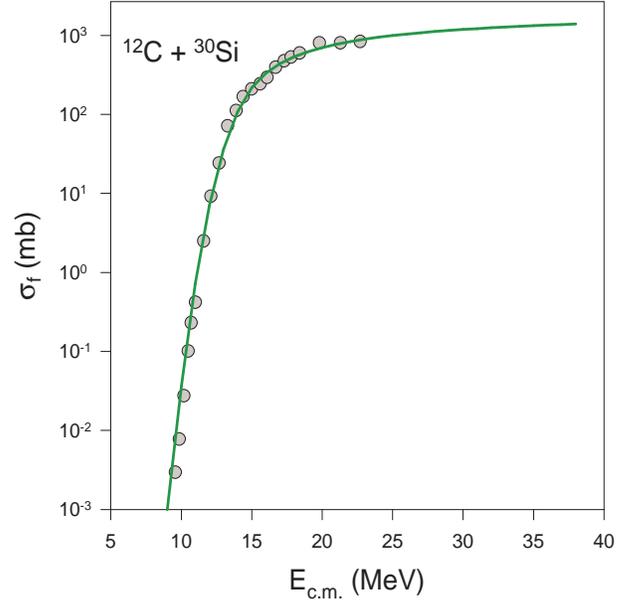,height=8cm,width=8cm}}
\caption
{Plots of the analytical estimates (solid line) and the measured values (full circles) of the capture excitation functions for $^{12}$C+$^{30}$Si.}
\label{fig2}
\vspace{0.0cm}
\end{figure}
\noindent 

\noindent
\section{Fusion cross section calculation}
\label{section3}

    In order to provide a systematic analysis of the data on the fusion excitation functions, a simple formula for the cross section for overcoming the potential energy barrier is derived. The energy dependence of the fusion cross section is obtained by folding the barrier distribution \cite{Wi04,Ca11} provided by Eq.(1), with the classical expression for the fusion cross section given by

\begin{eqnarray}
 \sigma_{f}(B) =&& \pi R_B^2 \Big[1-\frac{B}{E}\Big] ~~~~~~~~~~~~~~{\rm for} ~~B\leq E  \nonumber\\
 =&&0 ~~~~~~~~~~~~~~~~~~~~~~~~~~~~~~{\rm for}~~B\geq E
\label{seqn2}
\end{eqnarray}
\noindent
where $R_B$ denotes the relative distance corresponding to the position of the barrier approximately, which yields

\begin{eqnarray}
 &&\sigma_c(E) = \int_{E_0}^\infty \sigma_{f}(B) D(B)dB \nonumber\\
 &&= \int_{E_0}^{B_0} \sigma_{f}(B) D(B)dB + \int_{B_0}^E \sigma_{f}(B) D(B)dB  \\
 &&= \pi R_B^2\frac{\sigma_B}{E\sqrt{2\pi}}\Big[\xi\sqrt{\pi}\Big\{{\rm erf}(\xi)+{\rm erf}(\xi_0)\Big\}   
 +e^{-\xi^2} -e^{-\xi_0^2}\Big]  \nonumber
\label{seqn3}
\end{eqnarray}
\noindent
where $E_0=0$ for positive $Q$ value reactions and $E_0=Q$ for negative $Q$ value reactions, $Q$ value being the sum of the rest masses of fusing nuclei minus rest mass of the resultant fused nucleus, 

\begin{eqnarray}
 \xi &&= \frac{E-B_0}{\sigma_B\sqrt{2}} \nonumber\\
 \xi_0 &&= \frac{B_0-E_0}{\sigma_B\sqrt{2}} 
\label{seqn4}
\end{eqnarray}
\noindent
and erf($\xi$) is the Gaussian error integral of argument $\xi$. The parameters $B_0$ and $\sigma_B$ along with $R_B$ is to be determined by fitting Eq.(3) along with Eq.(4) to a given fusion excitation function. In the derivation of formula Eq.(3), the quantum effect of sub-barrier tunneling is not accounted for explicitly. However, the influence of the tunneling on shape of a given fusion excitation function is effectively included in the width parameter $\sigma_B$. 

    The fusion cross section formula of the Eq.(3) obtained by using the diffused-barrier, is a very elegant parametrization of the cross section for a process of overcoming the potential-energy barrier. Hence, it can be successfully used for analysis and predictions of the fusion excitation functions of light, medium and moderately heavy systems, especially in the range of sub-barrier energies. 

    In case of light and medium systems, surmounting the barrier automatically guarantees fusion of the colliding nuclei leading to formation of the compound nucleus. The term capture is used to refer the process of overcoming the interaction barrier in a nucleus-nucleus collision, followed by formation of a composite system. In general, the composite system undergoes fusion only in a fraction $f$ of the capture events. For light and medium systems, $f\approx1$, and almost all the capture events lead to fusion resulting fusion cross sections to be practically identical with the capture cross sections. However, for very heavy systems, only a small fraction ($f<1$) of capture events ultimately lead to fusion while for the remaining part of the events, the system re-separates prior to equilibration and clear distinction between fusion and capture cross sections then becomes necessary. Therefore, for very heavy systems, when the overcoming the barrier does not guarantee fusion, predictions based on Eq.(3) provide the capture cross section.

\begin{figure}[t]
\vspace{0.8cm}
\eject\centerline{\epsfig{file=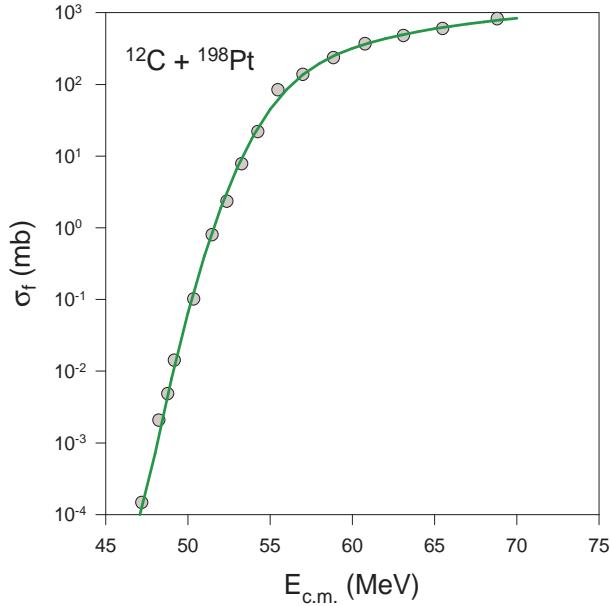,height=8cm,width=8cm}}
\caption
{Plots of the analytical estimates (solid line) and the measured values (full circles) of the capture excitation functions for $^{12}$C+$^{198}$Pt.}
\label{fig3}
\vspace{0.0cm}
\end{figure}
\noindent
  
\begin{figure}[t]
\vspace{0.8cm}
\eject\centerline{\epsfig{file=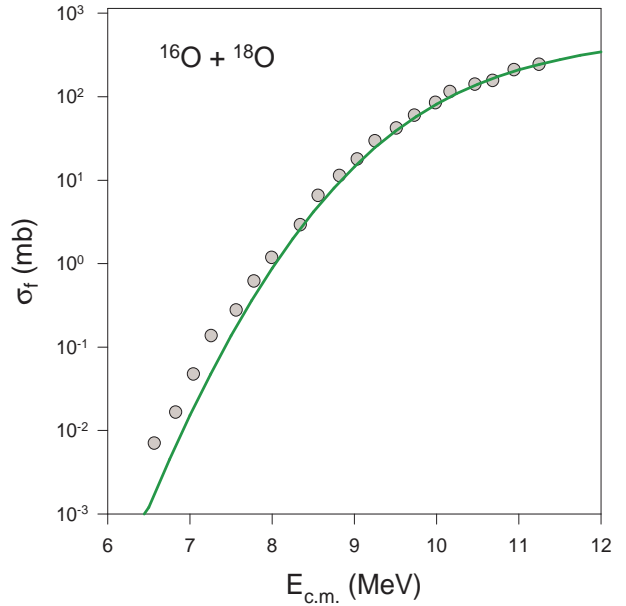,height=8cm,width=8cm}}
\caption
{Plots of the analytical estimates (solid line) and the measured values (full circles) of the capture excitation functions for $^{16}$O+$^{18}$O.}
\label{fig4}
\vspace{0.0cm}
\end{figure}
\noindent
        
\noindent
\section{ Results and discussion }
\label{section4}

    The sub-barrier fusion excitation functions of heavy-ion fusion for lighter systems of astrophysical interest have been analyzed using a simple diffused barrier formula (given by Eq.(1)) derived by folding the Gaussian barrier distribution with the classical expression for the fusion cross section for a fixed barrier. The same set of target-projectile combinations have been selected for which heavy ion sub-barrier fusion has been recently \cite{Mo97,Mo20,Ra20,We21} studied. The values of mean barrier height $B_0$, width $\sigma_B$ and the effective radius $R_B$ have been obtained using the least-square fit method. These values are listed in Table-I arranging it in order of increasing projectile mass.

\begin{figure}[t]
\vspace{0.8cm}
\eject\centerline{\epsfig{file=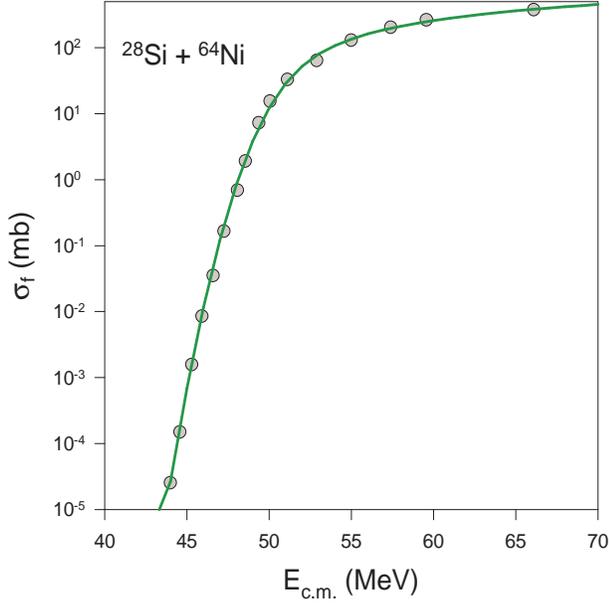,height=8cm,width=8cm}}
\caption
{Plots of the analytical estimates (solid line) and the measured values (full circles) of the capture excitation functions for $^{28}$Si+$^{64}$Ni.}
\label{fig5}
\vspace{0.0cm}
\end{figure}
\noindent 

\begin{figure}[t]
\vspace{0.8cm}
\eject\centerline{\epsfig{file=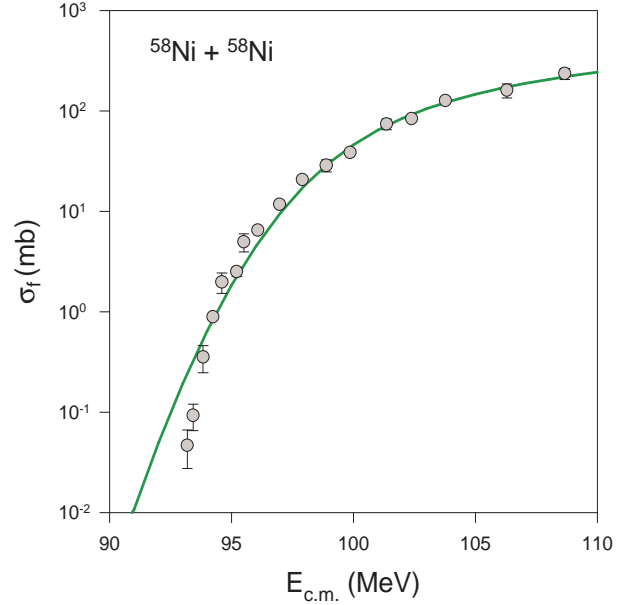,height=8cm,width=8cm}}
\caption
{Plots of the analytical estimates (solid line) and the measured values (full circles) of the capture excitation functions for $^{58}$Ni+$^{58}$Ni.}
\label{fig6}
\vspace{0.0cm}
\end{figure}
\noindent 

    The task of estimating $\sigma_c(E)$ rests upon predicting the values of the mean barrier height $B_0$, the width of the barrier height distribution $\sigma_B$ and the effective radius $R_B$ for a particular reaction. Since $B_0$ is essentially the mean height of the Coulomb barrier, it should be a function of the Coulomb parameter $z = Z_1Z_2/(A_1^{1/3} + A_2^{1/3})$ in the vicinity of the barrier. The second quantity, the effective barrier radius $R_B$, quite obviously should depend upon $r_0(A_1^{1/3} + A_2^{1/3}) = r_0 A_{12}$ where $r_0$ is the nuclear radius parameter. The extrapolation of the trend of $\sigma_B$ is more difficult, which primarily arises out of nuclear deformation, nuclear vibrations and quantum mechanical barrier penetrability. It may be observed from Table-I that the mean barrier height $B_0$ increases with the Coulomb parameter $z$ for all cases while the effective radius $R_B$ also increases with $A_{12}$ except for $^{16}$O+$^{18}$O and $^{28}$Si+$^{64}$Ni and interestingly, as may be seen from Figs.-1-7, for $^{16}$O+$^{18}$O case, the theoretical calculations are slightly off at lower energies as well.      

\begin{table}[h]
\vspace{0.0cm}
\caption{\label{tab:table1} The extracted values of the mean barrier height $B_0$, the width of the barrier height distribution $\sigma_B$ and the effective radius $R_B$, deduced from the analysis of the measured fusion excitation functions. The table is arranged in order of increasing projectile mass.}
\begin{tabular}{|c|c|c|c|c|c|c|}
\hline
Reaction &Refs.&$z$&$A_{12}$&$\sigma_B$&$B_0$&$R_B$ \\
& &  &  & [MeV] & [MeV] & [fm]         \\  
\hline

$^{12}$C+$^{24}$Mg&  \cite{Mo20}&13.916 &5.174  &0.815 &11.483 &6.646 \\   
$^{12}$C+$^{30}$Si&  \cite{Mo97}&15.565 &5.397  &1.090 &13.540 &8.300 \\  
$^{12}$C+$^{198}$Pt& \cite{Ra20}&57.650 &8.118  &1.749 &55.140 &11.179 \\ 
$^{16}$O+$^{18}$O&   \cite{Mo20}&12.450 &5.141  &0.859 &9.797 &7.743 \\
$^{28}$Si+$^{64}$Ni& \cite{We21}&55.709 &7.037  &1.402 &50.403 &7.182 \\
$^{58}$Ni+$^{58}$Ni& \cite{Ra20}&101.269&7.742  &2.275 &98.278 &8.550 \\ 
$^{64}$Ni+$^{64}$Ni& \cite{Ra20}&98.000 &8.000  &1.466 &92.646 &8.862 \\
\hline
\end{tabular} 
\vspace{0.0cm}
\end{table}

    In Figs.-1-7, the measured fusion excitation functions represented by full circles are compared with the predictions of the diffused barrier formula depicted by the solid lines. The systems of $^{12}$C+$^{24}$Mg, $^{12}$C+$^{30}$Si, $^{12}$C+$^{198}$Pt, $^{16}$O+$^{18}$O, $^{28}$Si+$^{64}$Ni, $^{58}$Ni+$^{58}$Ni and $^{64}$Ni+$^{64}$Ni are illustrated in Figs.-1-7. It may be easily perceived from these figures that precisely measured fusion excitation functions provide systematic information on the essential characteristics of the interaction potential, {\it viz.} the mean barrier width $\sigma_B$ and height $B_0$ of its distribution, for nucleus-nucleus collisions. The fusion or capture cross sections for planning experiments can also be predicted by using Eq.(3) and theoretically obtained values of the parameters $B_0$ and $\sigma_B$. 
    
    As seen in the Figs.-1-7, the present theoretical description provides excellent fits to the experimental data. This implies that for the chosen set of nuclei almost all the capture events lead to fusion resulting fusion cross sections to be practically identical with the capture cross sections. Moreover, the Gaussian form for the barrier distribution describes sub-barrier fusion cross sections quite well justifying the beyond single barrier model arising out of tunneling, deformation and vibration of nuclei. Although theoretically the concept of a barrier distribution is valid under certain approximations, the good fits to the experimental data shown certainly imply that in reactions involving heavy-ion fusion for lighter systems of astrophysical interest, barrier distribution remains a meaningful concept. 
    
\begin{figure}[t]
\vspace{0.8cm}
\eject\centerline{\epsfig{file=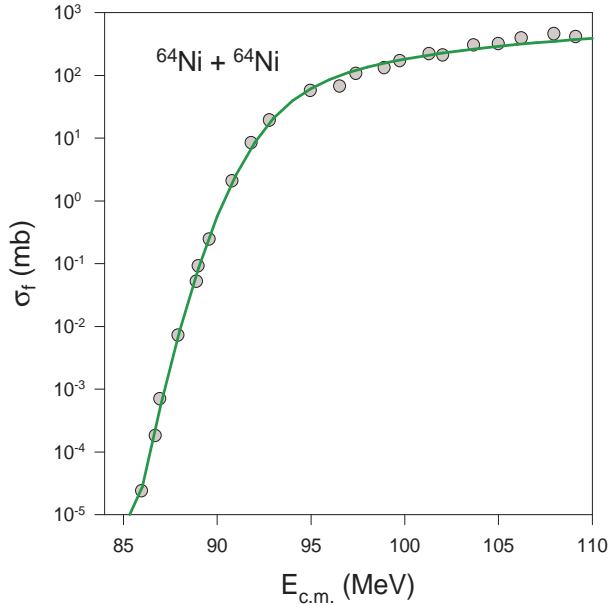,height=8cm,width=8cm}}
\caption
{Plots of the analytical estimates (solid line) and the measured values (full circles) of the capture excitation functions for $^{64}$Ni+$^{64}$Ni.}
\label{fig7}
\vspace{0.0cm}
\end{figure}
\noindent  

    It is pertinent to mention here that although the single-Gaussian parameterization [Eq.(1)] for barrier distribution is reasonably successful in providing a good description of fusion process in general, neither the formula derived for fusion cross-section nor the method using the barrier distribution can be put to use for all fusing systems. One can visualize from Eq.(3) that the excitation function is all the time a monotonically rising function of energy. This puts a limitation on Eqs.(1,2) which can not be used for describing fusion reactions at higher energies when incomplete fusion as well as deep-inelastic scattering can cause a lowering of the fusion cross section. Similar limitation arises for lighter systems (e.g. $^{12}$C+$^{12}$C, $^{12}$C+$^{16}$O, $^{16}$O+$^{16}$O etc.) as well when excitation functions possess oscillations and resonance structures. Possibility of better agreement with data may be explored by opting a more intricate formula for barrier distribution, which, however, will bring in more additional adjustable parameters than just three used in the present work. Such refinements for the barrier distribution of Eq.(1) include distributions with different widths on the low or high energy sides, a moderation of the exponent in the Gaussian distribution or multi-component distributions.

\noindent
\section{ Summary and conclusion }
\label{section5}

    In the region of sub-barrier energies, the fusion reaction cross sections have been estimated spanning a broad energy range. To comprehend the conditions of overcoming the potential energy barrier in nuclear collisions and to obtain a systematic knowledge on the essential characteristics of the interacting potential, {\it viz.} the mean barrier height $B_0$ and its width $\sigma_B$ of distribution, a set of precisely measured fusion excitation functions has been studied for two colliding nuclei. A Gaussian distribution function for the barrier heights is assumed to derive a simple diffused-barrier formula. The values of the mean barrier height $B_0$, the width $\sigma_B$ and the effective radius $R_B$ are determined using the method of least-square fit.          
  
    The present fusion cross section formula can be used to calculate the cross sections for overcoming the barrier in collisions of very heavy systems in order to calculate the overcoming-the-barrier cross section for a given projectile-target combination. For calculating the production cross sections of superheavy nuclei, the prediction of the capture excitation functions or sticking can be used in the sticking-diffusion-survival model \cite{An93} as one of three basic ingredients. The reasonably good fit of the present theoretical description to the experimental data implies two important facts that for the investigated set of nuclei almost all the capture events ultimately lead to fusion and the idea of the Gaussian distribution of barrier provides good description of the sub-barrier fusion cross sections. Although the single-Gaussian parameterization for barrier distribution is reasonably successful in providing a good description of fusion process in general, neither the formula derived for fusion cross-section nor the method using the barrier distribution can be put to use for all fusing systems. One can visualize from Eq.(3) that the excitation function is all the time a monotonically rising function of energy. This puts a limitation on Eqs.(1,2) which can not be used for describing fusion reactions at higher energies when incomplete fusion as well as deep-inelastic scattering can cause a lowering of the fusion cross section. Similar limitation can arise for lighter systems also when excitation functions possess oscillations and resonance structures. Possibility of better agreement with data may be explored by opting a more intricate formula for barrier distribution, which, however, will bring in more additional adjustable parameters than just three used in the present work. Such improvements for the barrier distribution can be realized through distributions with different widths on the low or high energy sides, multi-component distributions or a modification of the exponent in the Gaussian distribution.

\end{document}